\documentclass[10pt,conference]{IEEEtran}
\IEEEoverridecommandlockouts
\usepackage{cite}
\usepackage{amsmath,amssymb,amsfonts}
\usepackage{algorithmic}
\usepackage{graphicx}
\usepackage{textcomp}
\usepackage{xcolor}
\usepackage{todonotes}
\usepackage[inline]{enumitem}

\usepackage{balance}

\usepackage{tikz}

\def\BibTeX{{\rm B\kern-.05em{\sc i\kern-.025em b}\kern-.08em
    T\kern-.1667em\lower.7ex\hbox{E}\kern-.125emX}}
\begin{document}

\title{2HCDL: Holistic Human-Centered Development Lifecycle 
}

\author{\IEEEauthorblockN{Said Daoudagh}
\IEEEauthorblockA{\textit{CNR-ISTI}\\
Pisa, Italy \\
said.daoudagh@isti.cnr.it}
\and
\IEEEauthorblockN{Eda Marchetti}
\IEEEauthorblockA{\textit{CNR-ISTI}\\
Pisa, Italy \\
eda.marchetti@isti.cnr.it}
\and
\IEEEauthorblockN{Oum-El-Kheir Aktouf}
\IEEEauthorblockA{\textit{Univ. Grenoble Alpes, Grenoble INP}\\
LCIS, Valence, France \\
oum-el-kheir.aktouf@lcis.grenoble-inp.fr}
}

\maketitle

\begin{abstract} 
The recent events affecting global society continuously highlight the need to change the development lifecycle of complex systems by promoting human-centered solutions that increase awareness and ensure critical properties such as security, safety, trust, transparency, and privacy. This fast abstract introduces the Holistic Human-Centered Development Lifecycle (2HCDL) methodology focused on: (i) the enforcement of human values and properties and (ii) the mitigation and prevention of critical issues for more secure, safe, trustworthy, transparent, and private development processes. 
\end{abstract}


\section{Introduction}
\label{sec:introduction}
Ensuring trustworthiness and safe operation in complex systems with vulnerable hardware and software components is of paramount importance. Recent events, such as cyberattacks on critical infrastructure and global service disruptions, underscore the urgent need for innovative engineering approaches. This paper introduces the Holistic Human-Centered Development Lifecycle (2HCDL) methodology, which integrates target properties (e.g., security, safety, trust, transparency, and privacy), aligns with industrial needs, and focuses on stakeholders' requirements. The paper presents the Envisioned Objectives (EOs) of the 2HCDL methodology, its proposed solution, a prototype architecture, and future research directions.

\section{Envisioned Objectives}
\label{sec:objectives}
To address the challenges in complex system development, the 2HCDL methodology targets the following Envisioned Objectives (EOs):

\begin{enumerate}
    \item Holistic approach (EO1): Managing software, hardware, automation, electronics, and stakeholders' expertise through comprehensive solutions~\cite{nicoletti2023industrial, shin2021smartx}.

    \item Human-centered approach (EO2): Aligning development with social and ethical values, sustainability, and trustworthiness, and involving diverse stakeholders~\cite{kehrbusch2023digital, nicoletti2023industrial}.

    \item Modeling the behavior (EO3): Considering behavioral profiles of stakeholders in system modeling, implementation, validation, and prediction, utilizing AI, Digital Twins, crowdsourcing, and collaborative platforms~\cite{devOpsDT2022}.

    \item Integrated by-design approach (EO4): Incorporating target properties as specific principles from the early stages of development to prevent flaws, vulnerabilities, and cybersecurity issues~\cite{nicoletti2023industrial}.

    \item Self-adaptation and prediction (EO5): Employing self-adaptive methodologies for efficient component validation, reducing development costs, and predicting issues~\cite{casimiro2021self, weyns2022self}.

    \item Multidisciplinary approach (EO6): Utilizing various sources of knowledge, such as law, standards, technical specifications, and best practices, for requirements elicitation~\cite{devexp,hernandez2023requirements}.

    \item Quantitative and Qualitative proposal and solutions (EO7): Employing quantitative and qualitative analysis for risk management, testing, monitoring, and analyzing cybersecurity risks and violations, and integrating standards, metrics, and guidelines~\cite{VANLOOY2021103413,hernandez2023requirements}.

    \item Combining different Xs (EO8): Integrating and analyzing different target properties (Xs), such as security (Sec), privacy (Pri), transparency (Tra), lawfulness (Law), accountability (Acc), auditability (Aud), and certification (Cer), for achieving the required quality level~\cite{ref1SPCPSSurvey2017, shin2021smartx}.
\end{enumerate}

\section{The 2HC Dev-X-Ops Methodology}
\label{sec:proposedSolution}
\begin{figure*}[htbp]
\centerline{\includegraphics[width=0.65\textwidth]{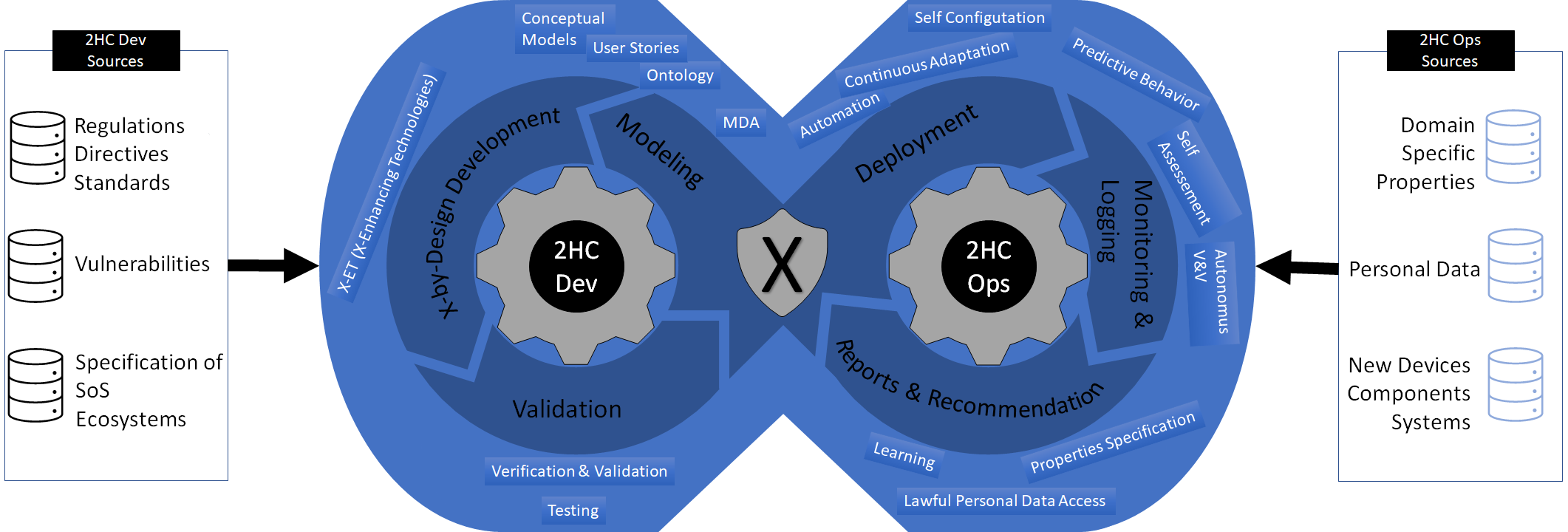}}
\caption{Holistic Human-Centered Dev-X-Ops (2HC Dev-X-Ops).}
\label{fig:DevXOps}
\end{figure*}

The 2HCDL methodology consists of two phases (see Figure~\ref{fig:DevXOps}): Holistic Human-Centered Development (2HC Dev) and Holistic Human-Centered Operation (2HC Ops). In the 2HC Dev phase, the methodology emphasizes modeling, X-by-Design development, and validation. It incorporates user research, user interface design, interaction design, accessibility, and usability testing. The 2HC Ops phase involves deployment, monitoring and logging, and reports \& recommendations, enabling self-assessment and prediction. The methodology supports continuous and incremental delivery, by-design principles, self-adaptation, and timely prediction.

\section{Architecture and Preliminary Implementation}
\label{sec:outcomes}
\begin{figure}[h!]
\centerline{\includegraphics[width=0.9\columnwidth]{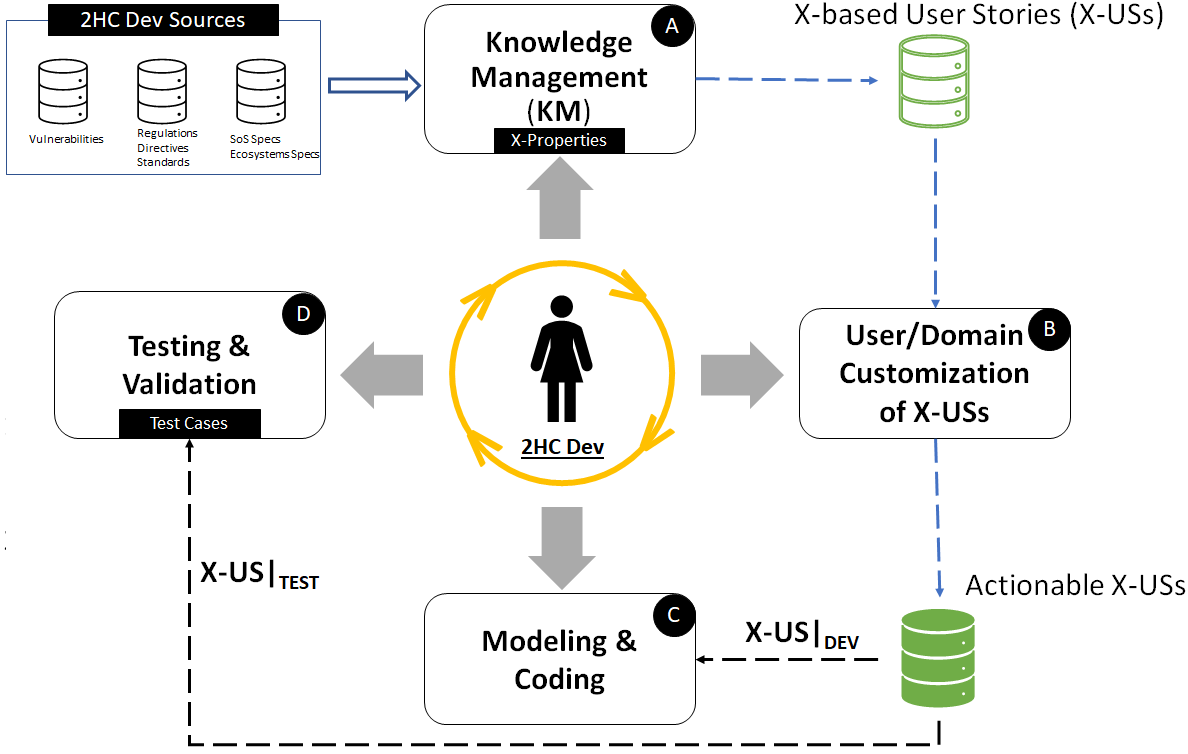}}
\caption{2HC Dev-X-Ops: Architecture Supporting Dev Phase.}
\label{fig:architecture}
\end{figure}

\begin{figure}[h!]
\centerline{\includegraphics[width=0.9\columnwidth]{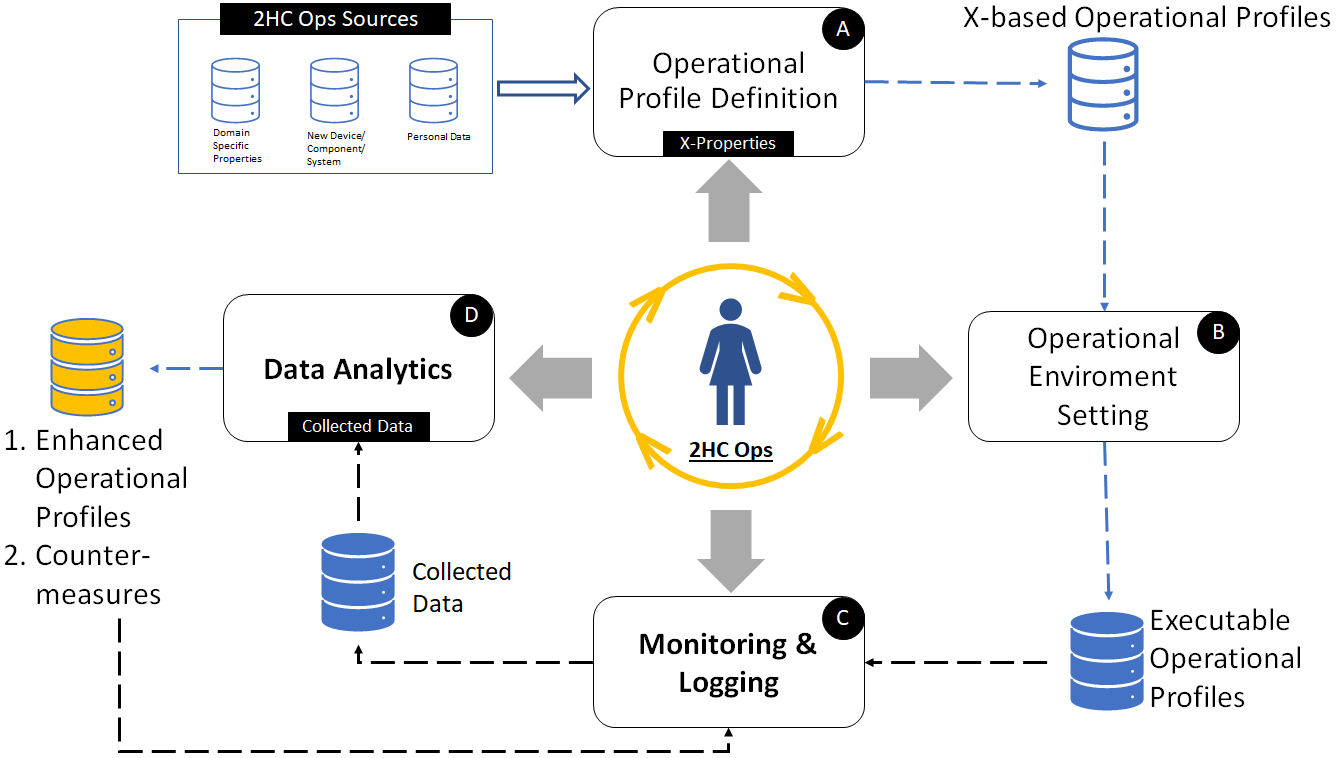}}
\caption{2HC Dev-X-Ops: Architecture Supporting Ops Phase.}
\label{fig:architecture2}
\end{figure}

The 2HCDL methodology is supported by a prototype architecture that accommodates the 2HC Dev (see Figure~\ref{fig:architecture}) and 2HC Ops (see Figure~\ref{fig:architecture2}) phases. The architecture includes components such as knowledge management, user/domain customization, modeling \& coding, testing \& validation, usage profile definition, operational environment setting, monitor \& logging, and data analytics. This architecture provides a reference for implementing 2HCDL and supports the development and operation of complex systems.

The 2HCDL methodology builds upon existing partial implementations such as DOXAT~\cite{DoxatDaoudaghLM23} and FIISS~\cite{FIISSPriyadarshiniGMA23}. DOXAT focuses on testing the Policy Decision Point (PDP) in access control systems, ensuring security and privacy. FIISS analyzes a target system's architectural and behavioral specifications to identify safety and security interactions. These implementations contribute to realising the 2HCDL methodology by integrating and extending them to cover other Xs properties.

\section{Conclusion}
\label{sec:conclusion}
The 2HCDL methodology offers a holistic, human-centered approach to system development, integrating critical properties and meeting stakeholders' requirements. By combining target properties, employing a multidisciplinary approach, and supporting self-adaptation and timely prediction, 2HCDL enables the development of trustworthy and safe systems. The prototype architecture serves as a foundation for implementation and further exploration.

\section*{Acknowledgment}
This work was partially supported by the French programme Projet ANR- 22-MRS0-0008-01 Programme MRSEIV3, the EU H2020 BIECO project Grant Agreement No. 952702 and the projects SERICS (PE00000014) and THE (CUP I53C22000780001) under the NRRP MUR program funded by the EU - NGEU.
\balance
\bibliographystyle{IEEEtran}
\bibliography{IEEEexample}

\end{document}